\documentclass[twocolumn,10pt]{asme2e}

\usepackage{titlesec}
\usepackage{graphicx}
\usepackage{amsmath}
\usepackage{float}
\usepackage{booktabs}

\titleformat{\section}
  {\normalfont\centering}
  {\Roman{section}.}{1em}{}
\titlespacing*{\section}{0pt}{12pt}{12pt}

\titleformat{\subsection}
  {\normalfont\itshape}
  {\Alph{subsection}.}{1em}{}
\titlespacing*{\subsection}{0pt}{10pt}{10pt}

\title{\underline{S}pherical \underline{Sail}ing \underline{O}mnidirectional \underline{R}over (SSailOR): Wind Tunnel Experimental Setup and Results}
\author{Aditya Varanwal
    \affiliation{
        \small Mechanical \& Aerospace Engineering Dept.\\
        \small North Carolina State University\\
        \small Raleigh, North Carolina 27695\\
        \small Email: avaranw@ncsu.edu
    }
}

\author{Parin Shah
    \affiliation{
        \small Mechanical \& Aerospace Engineering Dept.\\
        \small North Carolina State University\\
        \small Raleigh, North Carolina 27695
    }
}

\author{George Carrion
    \affiliation{
        \small Mechanical \& Aerospace Engineering Dept.\\
        \small North Carolina State University\\
        \small Raleigh, North Carolina 27695
    }
}

\author{Ashley Ortenburg
    \affiliation{
        \small Mechanical Engineering Dept.\\
        \small University of Michigan\\
        \small Ann Arbor, Michigan 48109
    }
}

\author{Diego Ramirez-Gomez
    \affiliation{
        \small Mechanical Engineering Dept.\\
        \small University of Michigan\\
        \small Ann Arbor, Michigan 48109
    }
}

\author{Chris Vermillion
    \affiliation{
        \small Mechanical Engineering Dept.\\
        \small University of Michigan\\
        \small Ann Arbor, Michigan 48109
    }
}

\author{Andre P. Mazzoleni
    \affiliation{
        \small Mechanical \& Aerospace Engineering Dept,\\
        \small North Carolina State University\\
        \small Raleigh, North Carolina 27695
    }
}

\begin{document}
\maketitle

\vspace{12pt}
\begin{abstract}

This paper presents the design, instrumentation, and experimental procedures used to test the Spherical Sailing Omnidirectional Rover (SSailOR) in a controlled wind tunnel environment. The SSailOR is a wind-powered autonomous rover. This concept is motivated by the growing need for persistent and sustainable robotic systems in applications such as planetary exploration, Arctic observation, and military surveillance. SSailOR uses wind propulsion via onboard sails to enable long-duration mobility with minimal energy consumption. The spherical design simplifies mechanical complexity while enabling omnidirectional movement. Experimental tests were conducted to validate dynamic models and assess the aerodynamic performance of the rover under various configurations and environmental conditions. As a result, this design requires a co-design approach. Details of the mechanical structure, sensor integration, electronics, data acquisition system, and test parameters are presented in this paper. In addition, key observations are made that are relevant to the design optimization for further development of the rover.
\end{abstract}

\vspace{12pt}
\section{INTRODUCTION}

\begin{figure}[H]
    \centering
    \includegraphics[width=0.95\linewidth]{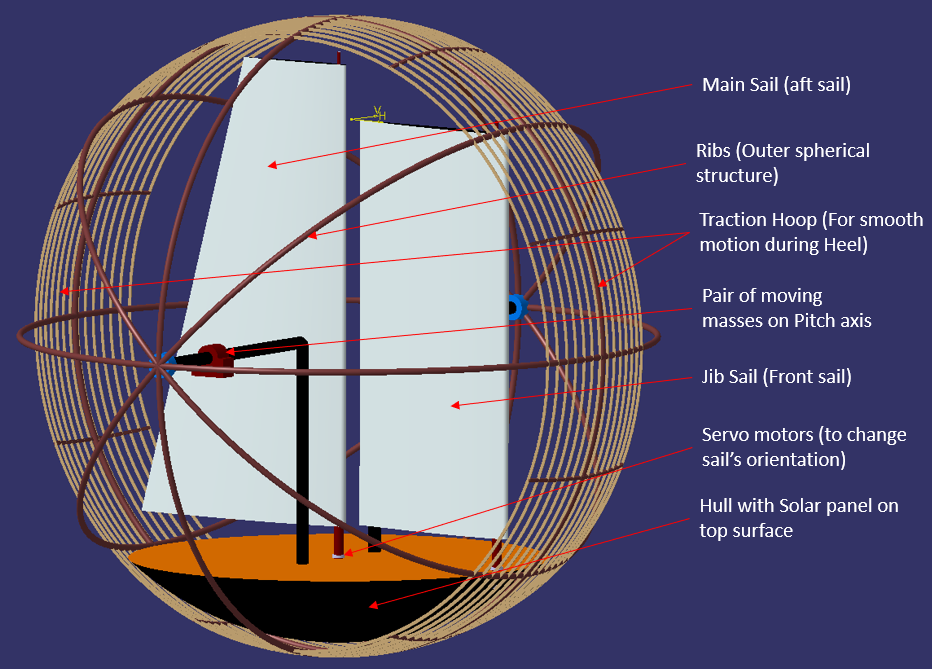}
    \caption{Schematic Diagram of SSailOR}
    \label{fig:ssailor_schematic}
\end{figure}

The Spherical Sailing Omnidirectional Rover (SSailOR) is a novel, wind-powered, spherical robotic platform designed for omnidirectional mobility (Figure~\ref{fig:ssailor_schematic}). This class of mobile systems is particularly relevant for missions that require persistent, sustainable, and autonomous travel in challenging or remote environments, such as Arctic exploration, military reconnaissance, or planetary surface missions.

SSailOR operates via two sails, namely a main sail and a jib sail, which are mounted on an inner hull. By leveraging environmental wind forces and adjusting sail angles, the rover achieves directional control and propulsion without active wheel-based drive systems. The system design is further enhanced by internal actuation mechanisms such as stepper motors and moving mass, which influence roll and yaw — making SSailOR a hybrid of passive and active mobility paradigms.

A simplified dynamic model for SSailOR has already been developed, incorporating basic aerodynamic interactions and internal actuation. However, for deployment and accurate trajectory prediction, this model must be validated and refined using real-world aerodynamic data.

To address this need, wind tunnel experiments were designed and conducted at North Carolina State University (NC State University). The subsonic wind tunnel at North Carolina State University was used due to its capability to produce repeatable, controlled freestream conditions. These experiments were aimed at:
a. Characterizing lift and drag forces under varying sail angles and rover yaw angles,
b. Validating the assumptions made in the simplified dynamic model,
c. Optimizing sail geometry for efficient propulsion,
d. Developing data-driven insights into subsystem-level and full-system aerodynamic performance.

\begin{table}[H]
\footnotesize
\caption{Major difference between other wind-driven rovers \& SSailOR}
\label{tab:individual_test_params}
\centering
\begin{tabular}{|l|l|l|}
\hline
\textbf{} & \textbf{Self-Righting} & \textbf{Upwind-Capable}\\ \hline
Land Sailor & Not Capable & Capable \\ \hline
Tumbleweed Rover &Capable &Not Capable \\ \hline
SSailOR &Capable & Capable \\ \hline
\end{tabular}
\end{table}

The experimental setup included a modular test rig with upper and lower mounts, precision sensors (including a six-axis load cell), stepper motor for rover rotation, and fairings to minimize extraneous wind effects. The test matrix covered a range of wind pressures, rotational speeds, yaw angles, and sail configurations, enabling comprehensive exploration of SSailOR's aerodynamic behavior in a controlled environment.

The results presented in this paper lay the groundwork for refined modeling, simulation-based optimization, and future field testing of SSailOR. They also contribute to a broader understanding of wind-driven spherical robotic systems as a sustainable solution for long-term autonomous exploration.

\vspace{12pt}
\section{\centering SYSTEM DESIGN}

\textbf{Design Expectations:}
\begin{enumerate}
  \item \textbf{Data Collection:} Quantify aerodynamic forces (lift, drag) across multiple configurations.
  \item \textbf{Mobility:} Ensure behavior mimics field-like motion.
  \item \textbf{Configurability:} Allow yaw variation and sail adjustment across a range of angles.
  \item \textbf{Isolation:} Design test mount to minimize external aerodynamic interference.
\end{enumerate}

\begin{figure}[H]
    \centering
    \includegraphics[width=0.95\linewidth]{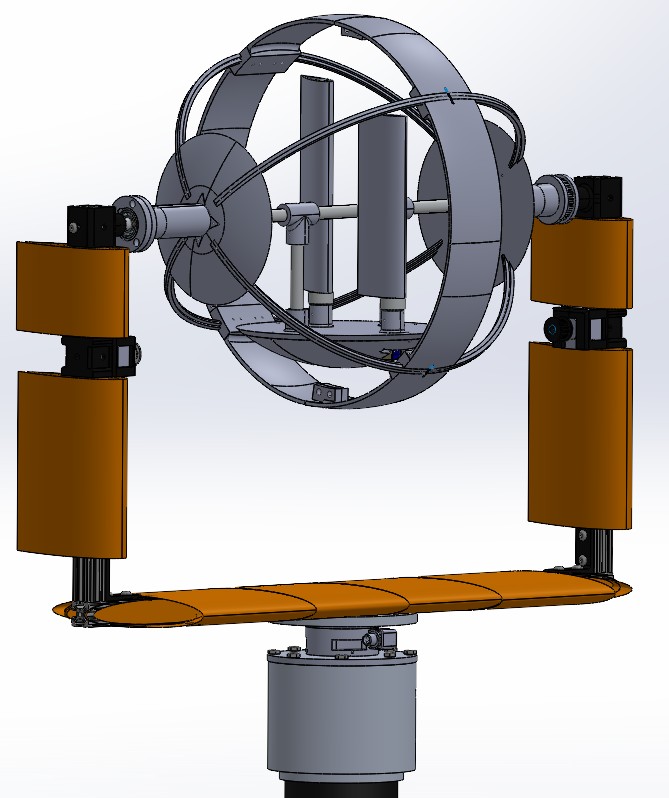}
    \caption{SSailOR Test Setup (CAD Model)}
    \label{fig:Test Setup}
\end{figure}

\textbf{Test Structure:} \\
The rig consists of the rover, an upper test mount, and a lower mount (Figure~\ref{fig:Test Setup}). The frame was designed to be modular for easy reconfiguration. Symmetry was maintained across the upper structure using vertical fairings and a dummy motor for structure symmetry. The entire setup was constrained to fit within the wind tunnel test section of 32x45x46 inches. So, considering the wind tunnel cross section blockage percentage, half scaled model (40 cm diameter SSailOR rover) has been developed.

To minimize flow disturbances:
\begin{itemize}
    \item An 80 mm clearance was maintained between the rover and support hardware.
    \item Vertical and horizontal symmetric airfoil and elliptical fairings respectively were added to streamline the support frame.
    \item Minimum wall and floor clearances were maintained as per wind tunnel best practices.
\end{itemize}

NACA 0030 and elliptical fairing shapes were selected based on geometry, expected flow speed, and wake effects. Selection rationale can be further backed using pressure drag calculations or Reynolds number analysis.

\textbf{Rover Design:} \\
The rover includes a traction hoop, curved ribs, side hubs, sails, and an inner hull. Ribs connect the hoop to side hubs via a curved T-frame structure to improve stiffness. The main sail and jib sail are designed using symmetric airfoils NACA0020 and NACA0025 airfoils respectively (Table~\ref{tab:main_test_params}).

\textbf{Data Collection:} \\
A six-axis load cell, ATI Gamma FT16284 model, was used to measure forces and moments. It was placed between the upper and lower mounts for accurate torque measurement. Wind speed and temperature data were acquired from the wind tunnel sensors via LabVIEW, and a separate barometric sensor recorded ambient pressure.

\begin{table}[H]
\footnotesize
\caption{System Design Parameters}
\label{tab:main_test_params}
\centering
\begin{tabular}{|l|l|}
\hline
\textbf{Parameter} & \textbf{Values} \\ \hline
Rover Diameter & 400 mm \\ \hline
Side Plate & 19.3 mm height and 174.4 mm dia\\ \hline
Main Sail Chord Length & 100 mm \\ \hline
Main Sail Span & 215 mm \\ \hline
Jib Sail Chord Length & 70 mm \\ \hline
Jib Sail Span & 185 mm \\ \hline
Hull Diameter & 215 mm \\ \hline
Ribs dimesnion & T cross section of 8 mm width \\ \hline
Number of Ribs & 8 nos. \\ \hline
Main Sail Angle (degree) & -8,-4,0,4,8,12,16 \\ \hline
\end{tabular}
\end{table}

\vspace{12pt}
\section{\centering MECHANICAL SETUP}

\textbf{Sensor Placement:} \\
The load cell is mounted beneath the test rig, between the test frame and the lower motor mount, ensuring accurate force and torque measurement. The sphere’s rotation is verified using a Hall-effect magnetic sensor mounted opposite the drive assembly. This sensor reads a rotating magnet on a cylindrical extension connected to the rover frame.

\textbf{Mobility:} \\
All rover components are fabricated using FDM 3D printing. Assemblies use standardized T-slotted rails, nut and bolt fasteners, M2 screws, for modularity. The hoop is assembled from multiple arc segments joined by interlocking 3-D printed sleeves. The spherical body is supported by a central aluminum rod with four bearings to allow low-friction rotation. Drive torque is applied via a stepper motor and timing belt connected to a driven gear. The selected gear ratio was 2:1.

\textbf{Frame:} \\
The main structural frame includes T-slot rails, a central aluminum rod, and 3D-printed connectors to ensure mechanical isolation of the rover from external disturbances. Vertical fairings rotate with the yaw axis and are fixed in position using mechanical collars. Horizontal fairings are directly attached to the lower frame. Symmetry was maintained by installing a second stepper motor on the opposing side (not powered during tests).

\begin{figure}[H]
    \centering
    \includegraphics[width=0.95\linewidth]{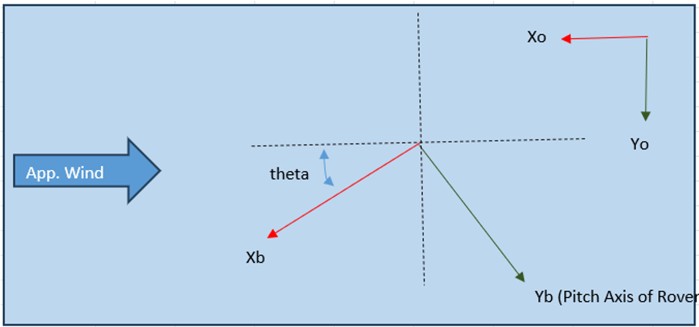}
    \caption{Reference Frame and Rover Orientation}
    \label{fig:ref_frame}
\end{figure}
\vspace{12pt}
\section{\centering ELECTRONICS SETUP}
The electronics were designed for synchronized actuation, sensing, and data acquisition. A micro-stepping driver controlled the stepper motor's speed and position via external commands, enabling precise rotation of the spherical body. The RPM was determined using a Hall-effect sensor that generated a pulse per revolution.

An upstream wind sensor, calibrated for the test section, measured free-stream velocity. The yaw orientation was tracked using an IMU, which provided angular feedback to support closed-loop control of the jib sail.

A servo motor actuated the jib sail based on real-time torque readings from the load cell. This allowed the system to actively maintain near-zero net torque about the vertical axis under varying flow conditions. The six-axis load cell simultaneously recorded forces ($F_x$, $F_y$, $F_z$) and moments ($M_x$, $M_y$, $M_z$).
\begin{figure}[H]
    \centering
    \includegraphics[width=0.95\linewidth]{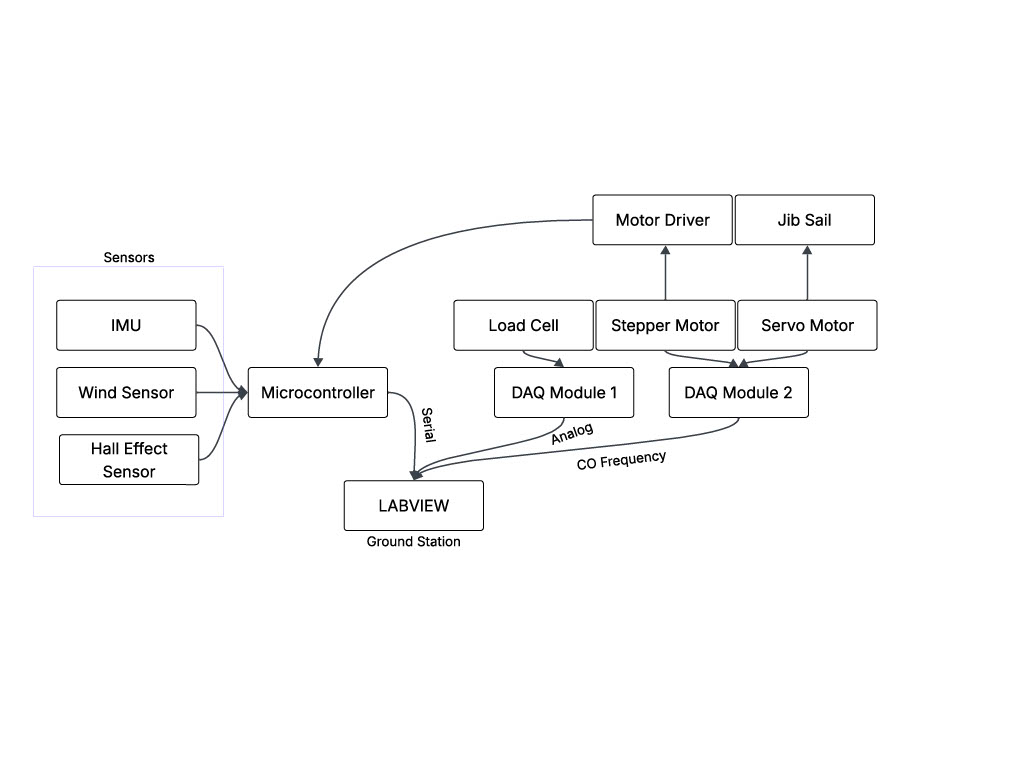}
    \caption{Block Diagram representing SSailOR's Electronics and DAQ}
    \label{fig:Electronics_BD}
\end{figure}

\vspace{12pt}
\section{\centering DATA ACQUISITION}

Sensor outputs were routed through two acquisition paths:
\begin{itemize}
    \item A microcontroller collected IMU, wind sensor, and Hall-effect data via digital input. It handled time stamping, basic processing, and forwarded data over serial to LabVIEW.
    \item Two National Instruments DAQ modules interfaced with analog channels. One module acquired six-axis load cell signals; the other generated PWM signals for stepper and servo motor control.
\end{itemize}

LabVIEW acted as the ground station GUI, providing real-time visualization of sensor streams, system health, and actuator state(Figure \ref{fig:Electronics_BD}). The integrated environment enabled feedback-controlled experiments with minimal latency. Data were logged at 100 Hz for all channels, ensuring high-resolution capture of unsteady aerodynamic events.

\vspace{12pt}
\section{\centering TEST DESIGN}

\begin{figure}[H]
    \centering
    \includegraphics[width=0.95\linewidth]{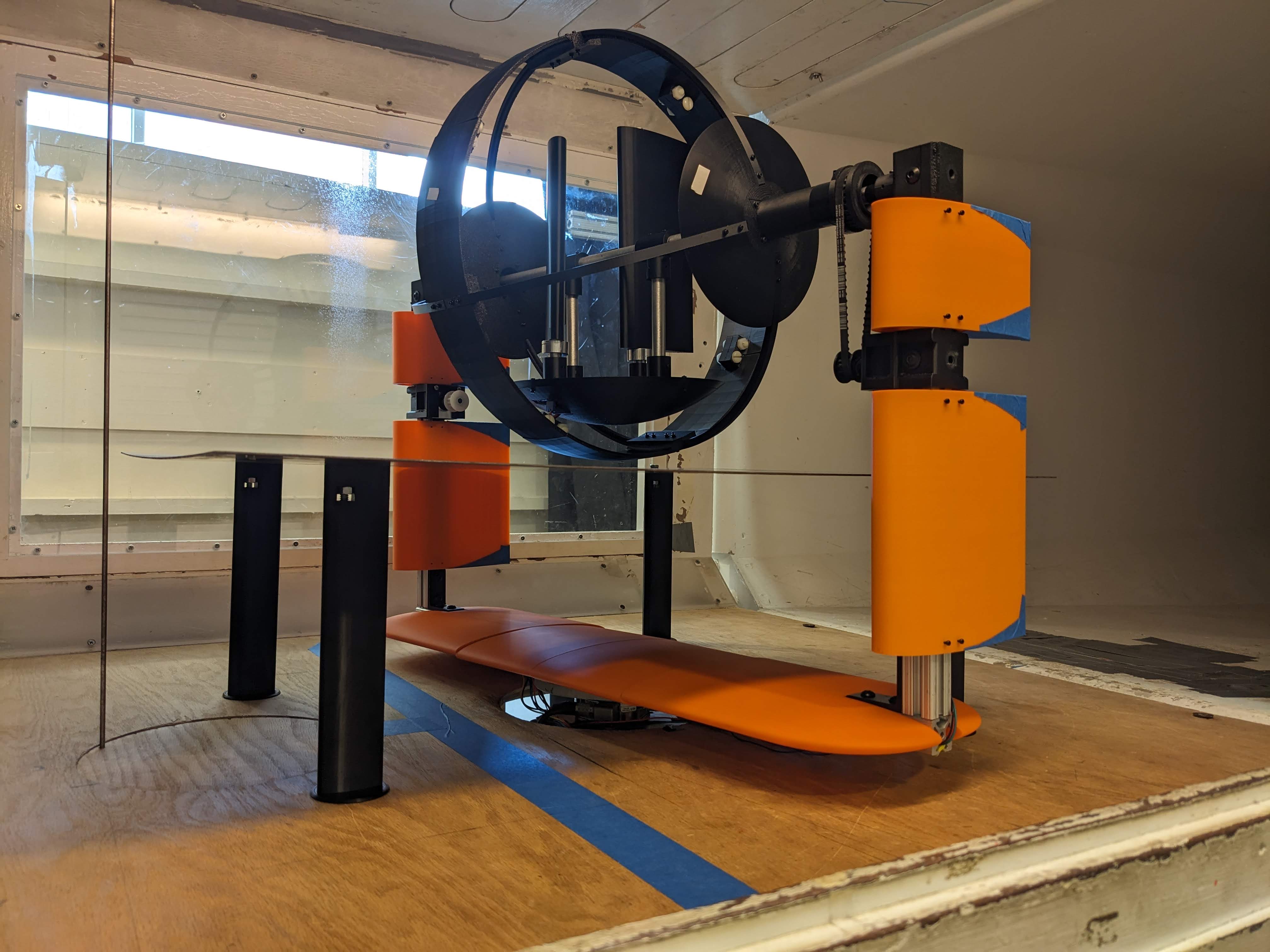}
    \caption{Ground Effect Test Configuration}
    \label{fig:ground test}
\end{figure}

\textbf{Baseline Testing:} \\
Full-system tests were conducted under various conditions to measure the aerodynamic forces and moments as functions of wind speed, yaw angle, sail configuration, and rotational state. In addition, a ground effect test was performed for the rover configuration by placing a flat steel plate beneath the SSailOR model(Figure \ref{fig:ground test}). This plate was supported by four NACA 0025 airfoils, allowing the evaluation of how proximity to the ground influences the measured aerodynamic forces. These main test parameters form the basis of SSailOR's performance evaluation in representative operating conditions. 

\begin{table}[H]
\footnotesize
\caption{Main Test Parameters}
\label{tab:main_test_params}
\centering
\begin{tabular}{|l|l|}
\hline 
\textbf{Parameter} & \textbf{Values} \\ \hline
RPM & 0, 60, 80 \\ \hline
PSF & 1.0, 1.5 \\ \hline
Yaw [$^\circ$] & 10, 20, 30, 40 \\ \hline
Setup Type & Constant RPM, Variable RPM \\ \hline
Main Sail Angle of Attack [$^\circ$] & 0, 4, 8, 12, 16 \\ \hline
\end{tabular}
\end{table}

\textbf{Subsystem Testing:} \\
To isolate aerodynamic contributions of individual components, tests were repeated under three configurations: structure frame only, hoop+ribs, and sails+inner-hub.

\begin{table}[H]
\footnotesize
\caption{Component-Level Test Configurations}
\label{tab:individual_test_params}
\centering
\begin{tabular}{|l|l|l|l|l|}
\hline
\textbf{Case} & \textbf{RPM} & \textbf{PSF} & \textbf{Yaw} & \textbf{Main Sail AOA} \\ \hline
Structure Only & N/A & 1.0, 1.5 & 10,20,30,40 & N/A \\ \hline
Ground Effect Test &60, 80 &1.0, 1.5 &10 &4, 8 \\ \hline
Hoop + Ribs & 0, 80 & 1.5 & 20,30,40 & N/A \\ \hline
Sails + Hub & N/A & 1.5 & 20,30,40 & 0,4,8,12 \\ \hline
\end{tabular}
\end{table}

\textbf{Geometric Specifications:}

\begin{figure}[H]
    \centering
    \includegraphics[width=0.95\linewidth]{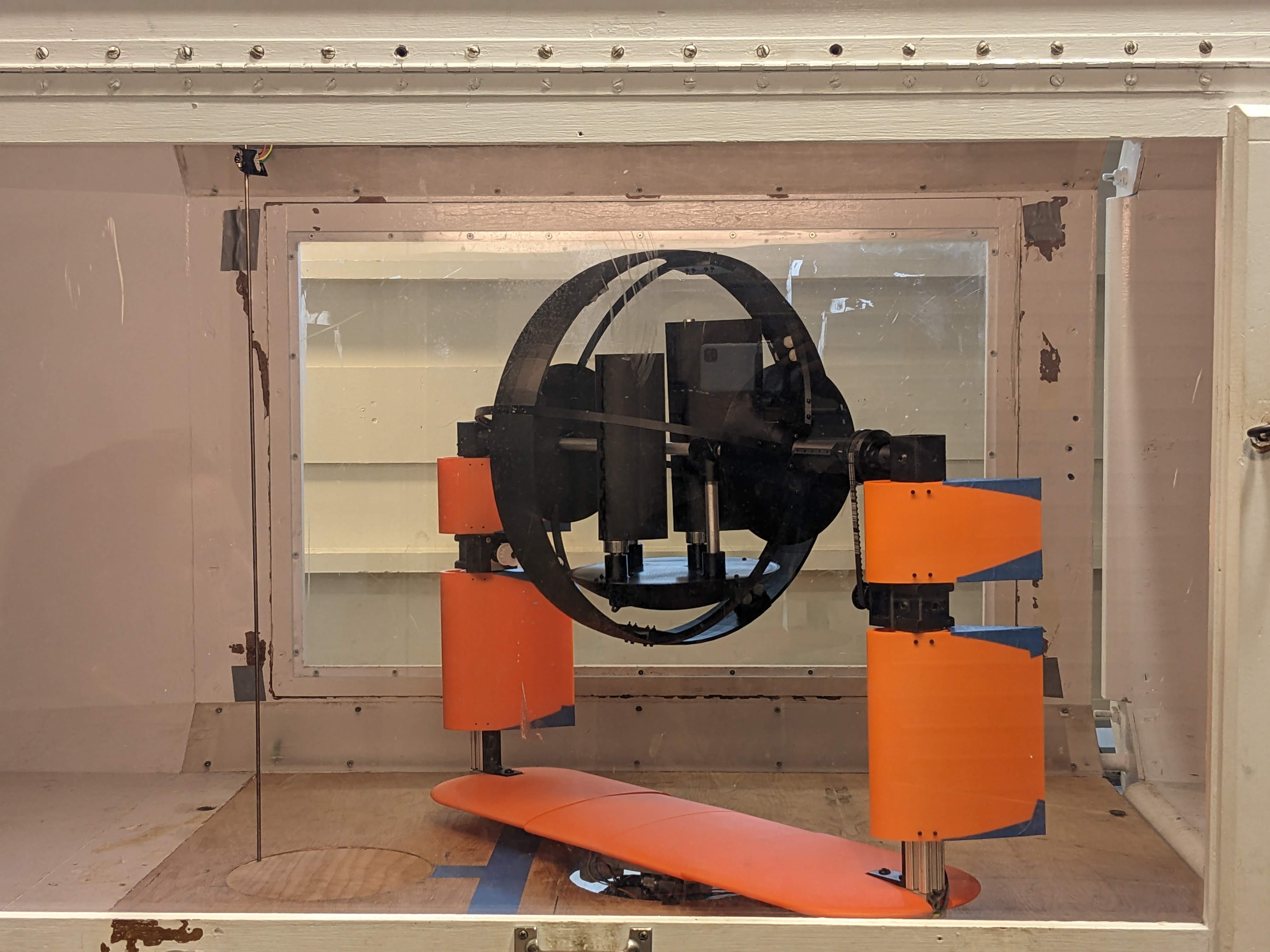}
    \caption{Complete Rover Test Configuration}
    \label{fig:rover_model}
\end{figure}

\begin{figure}[H]
    \centering
    \includegraphics[width=0.95\linewidth]{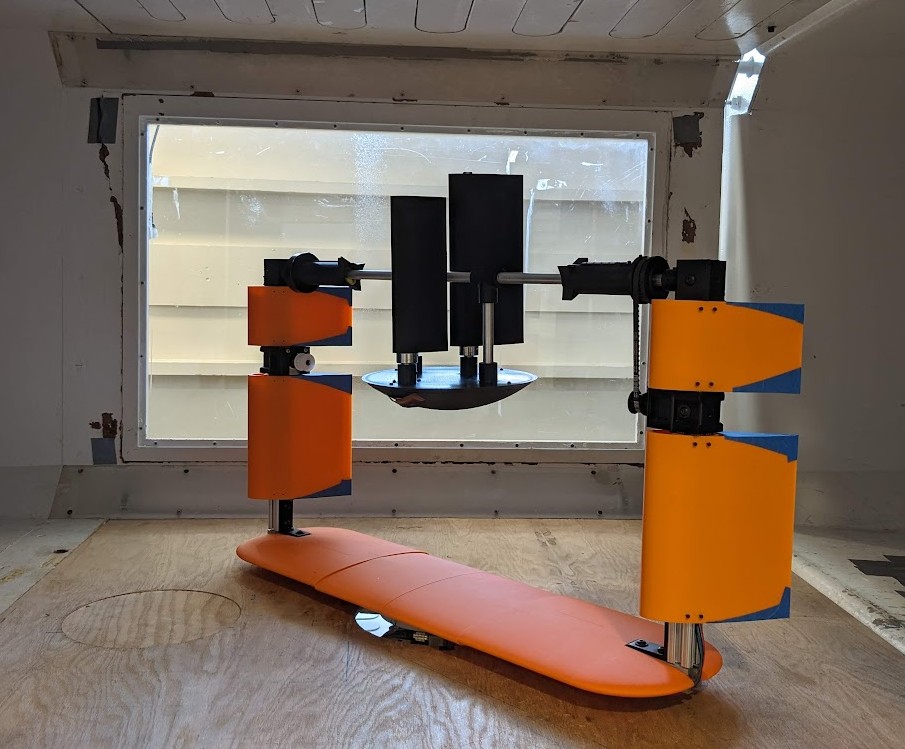}
    \caption{Only Sails COnfiguration Test}
    \label{fig:onlysails}
\end{figure}

\begin{figure}[H]
    \centering
    \includegraphics[width=0.95\linewidth]{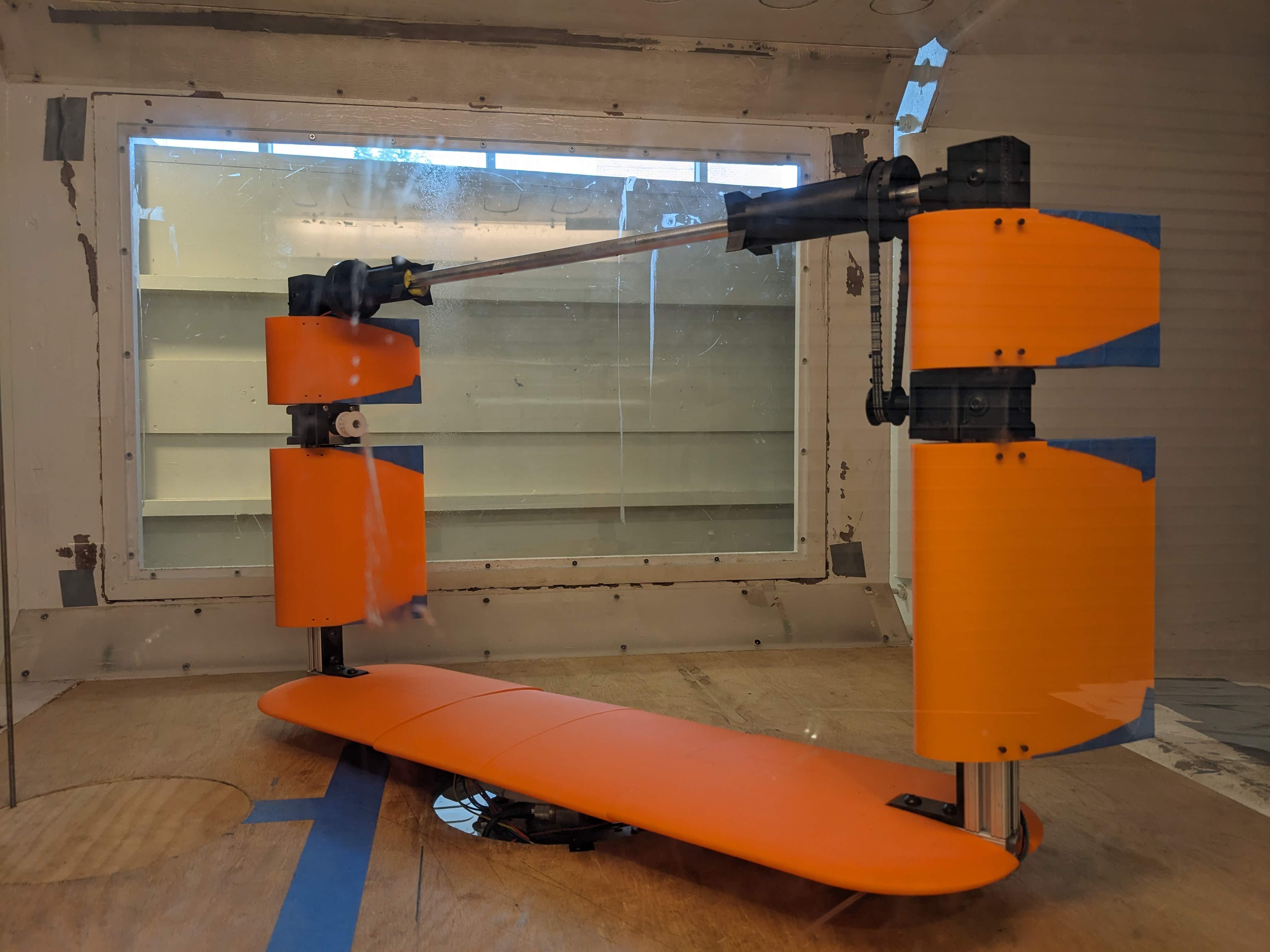}
    \caption{Structure Test Configuration}
    \label{fig:structure test}
\end{figure}

\vspace{12pt}
\section{\centering RESULTS}

\begin{figure}[H]
    \centering
    \includegraphics[width=0.95\linewidth]{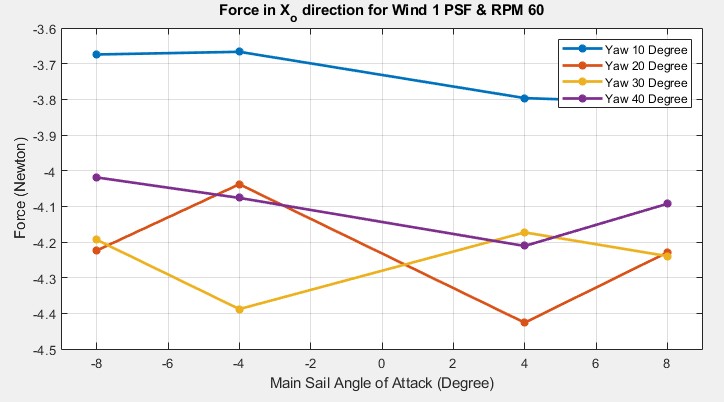}
    \caption{Longitudinal force (i.e drag) data for varying Yaw at 1 PSF 60 RPM}
    \label{fig:results_fx1}
\end{figure}

\begin{figure}[H]
    \centering
    \includegraphics[width=0.95\linewidth]{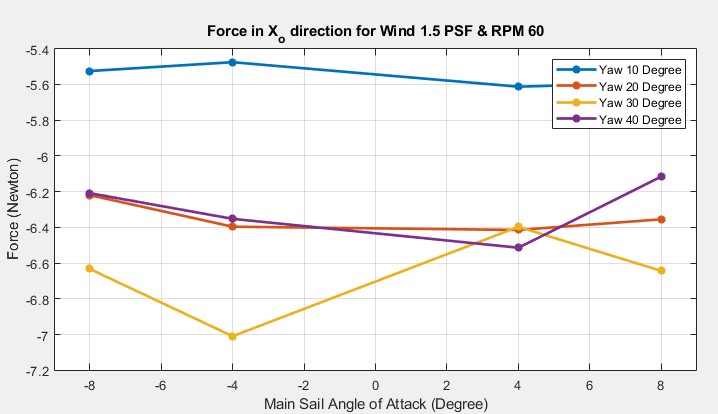}
    \caption{Longitudinal force (i.e drag) data for varying Yaw at 1.5 PSF 60 RPM}
    \label{fig:results_fx2}
\end{figure}

\begin{figure}[H]
    \centering
    \includegraphics[width=0.95\linewidth]{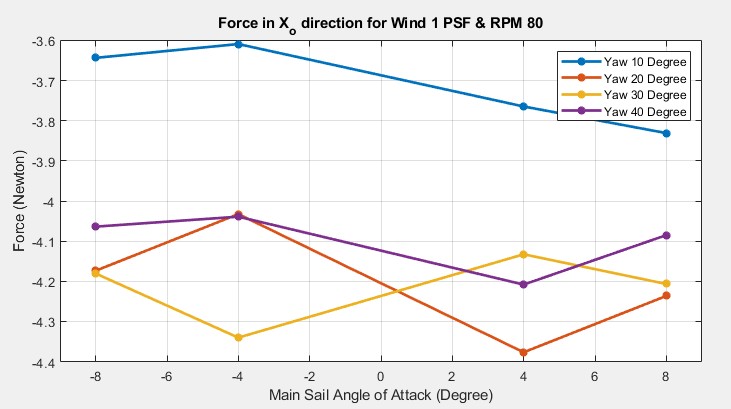}
    \caption{Longitudinal force (i.e drag) data for varying Yaw at 1 PSF 80 RPM}
    \label{fig:results_fx3}
\end{figure}

\begin{figure}[H]
    \centering
    \includegraphics[width=0.95\linewidth]{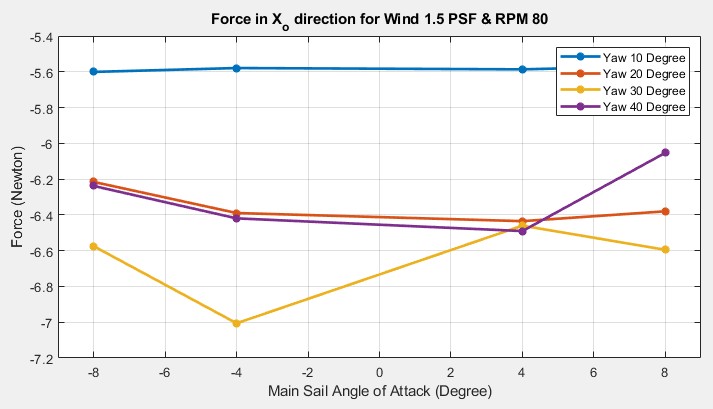}
    \caption{Longitudinal force (i.e drag) data for varying Yaw at 1.5 PSF 80 RPM}
    \label{fig:results_fx4}
\end{figure}

\begin{figure}[H]
    \centering
    \includegraphics[width=0.95\linewidth]{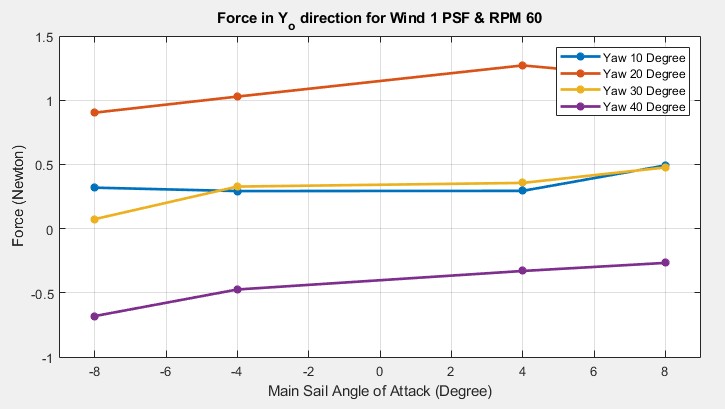}
    \caption{Longitudinal force (i.e drag) data for varying Yaw at 1 PSF 60 RPM}
    \label{fig:results_fy1}
\end{figure}

\begin{figure}[H]
    \centering
    \includegraphics[width=0.95\linewidth]{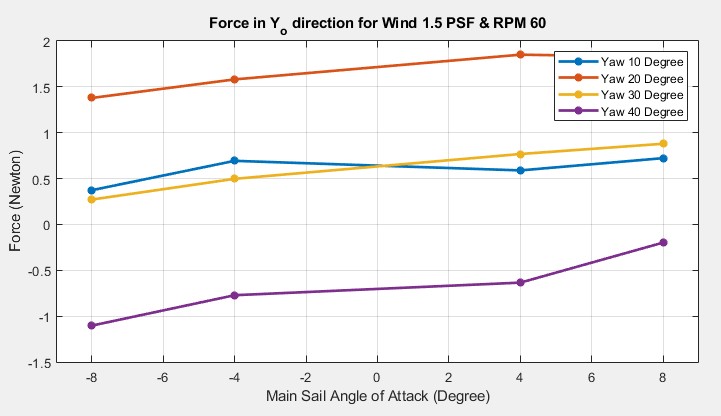}
    \caption{Longitudinal force (i.e drag) data for varying Yaw at 1.5 PSF 60 RPM}
    \label{fig:results_fy2}
\end{figure}

\begin{figure}[H]
    \centering
    \includegraphics[width=0.95\linewidth]{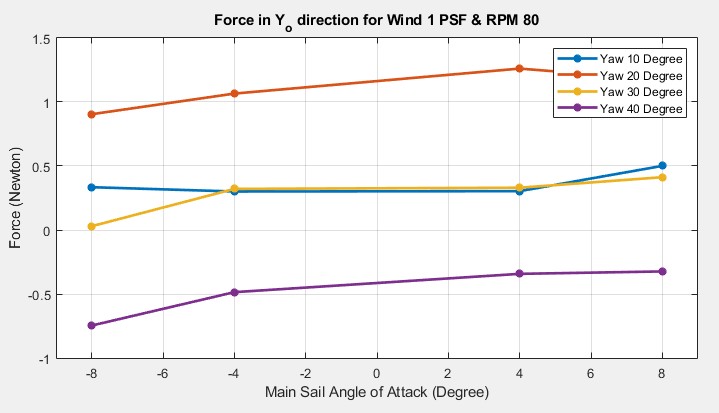}
    \caption{Longitudinal force (i.e drag) data for varying Yaw at 1 PSF 80 RPM}
    \label{fig:results_fy3}
\end{figure}

\begin{figure}[H]
    \centering
    \includegraphics[width=0.95\linewidth]{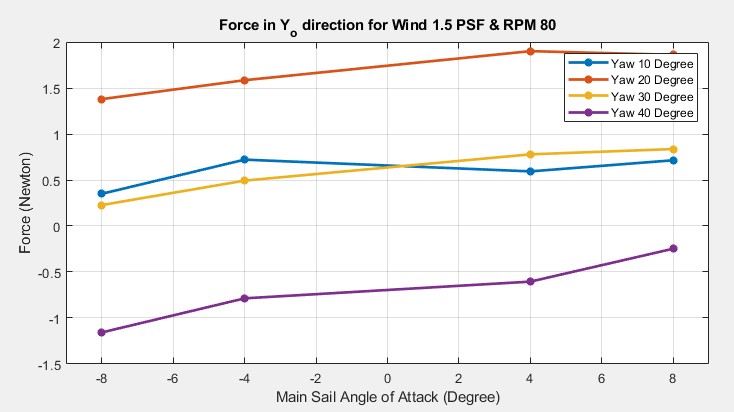}
    \caption{Lateral force (i.e Lift) data for varying Yaw at 1.5 PSF 80 RPM}
    \label{fig:results_fy4}
\end{figure}

\begin{figure}[H]
    \centering
    \includegraphics[width=0.95\linewidth]{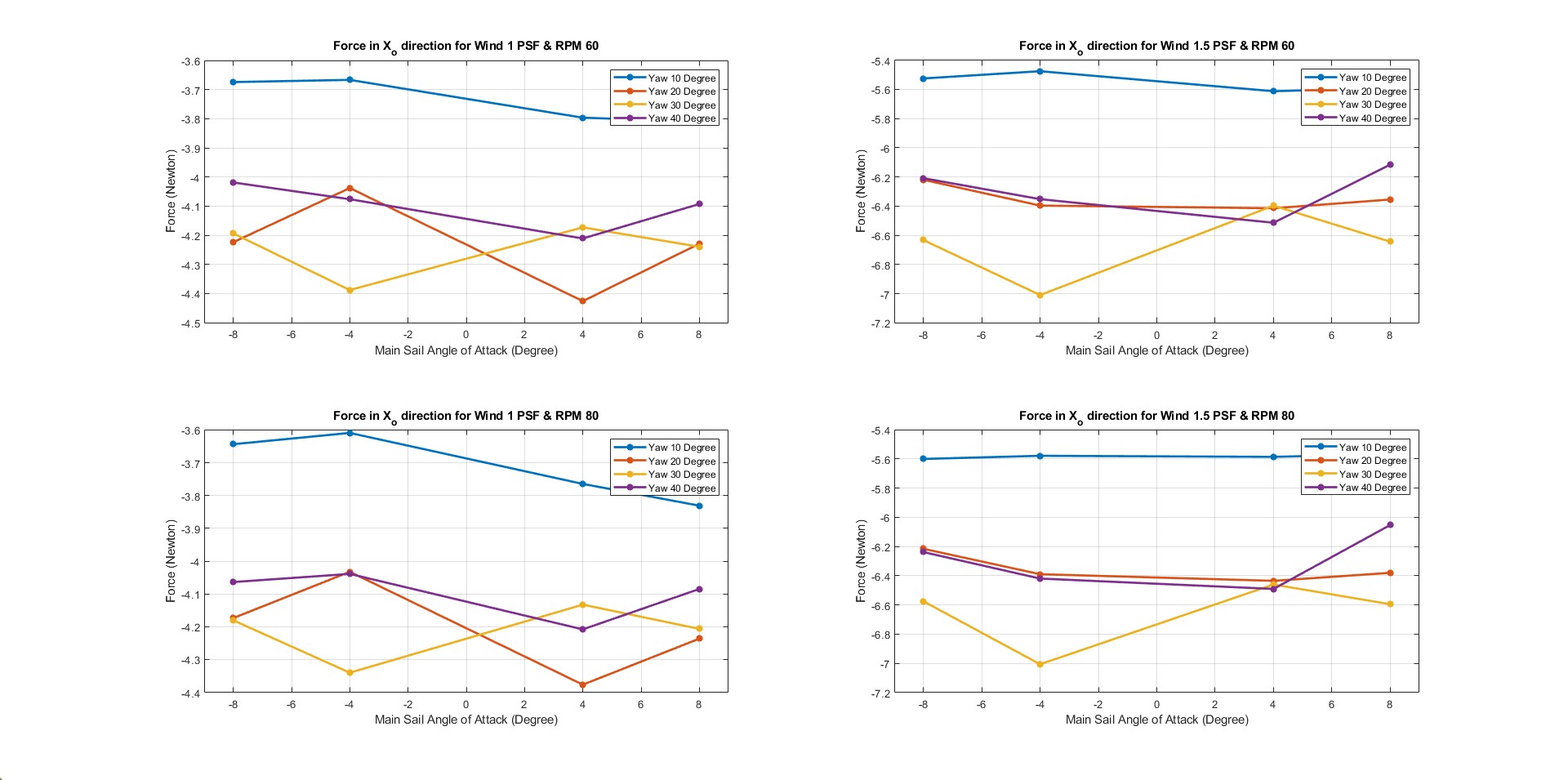}
    \caption{Compiled Plot: Longitudnal ($F_x$) force (i.e. Lift) data for main sail angle sweep.}
    \label{fig:results_compliled_fx}
\end{figure}

\begin{figure}[H]
    \centering
    \includegraphics[width=0.95\linewidth]{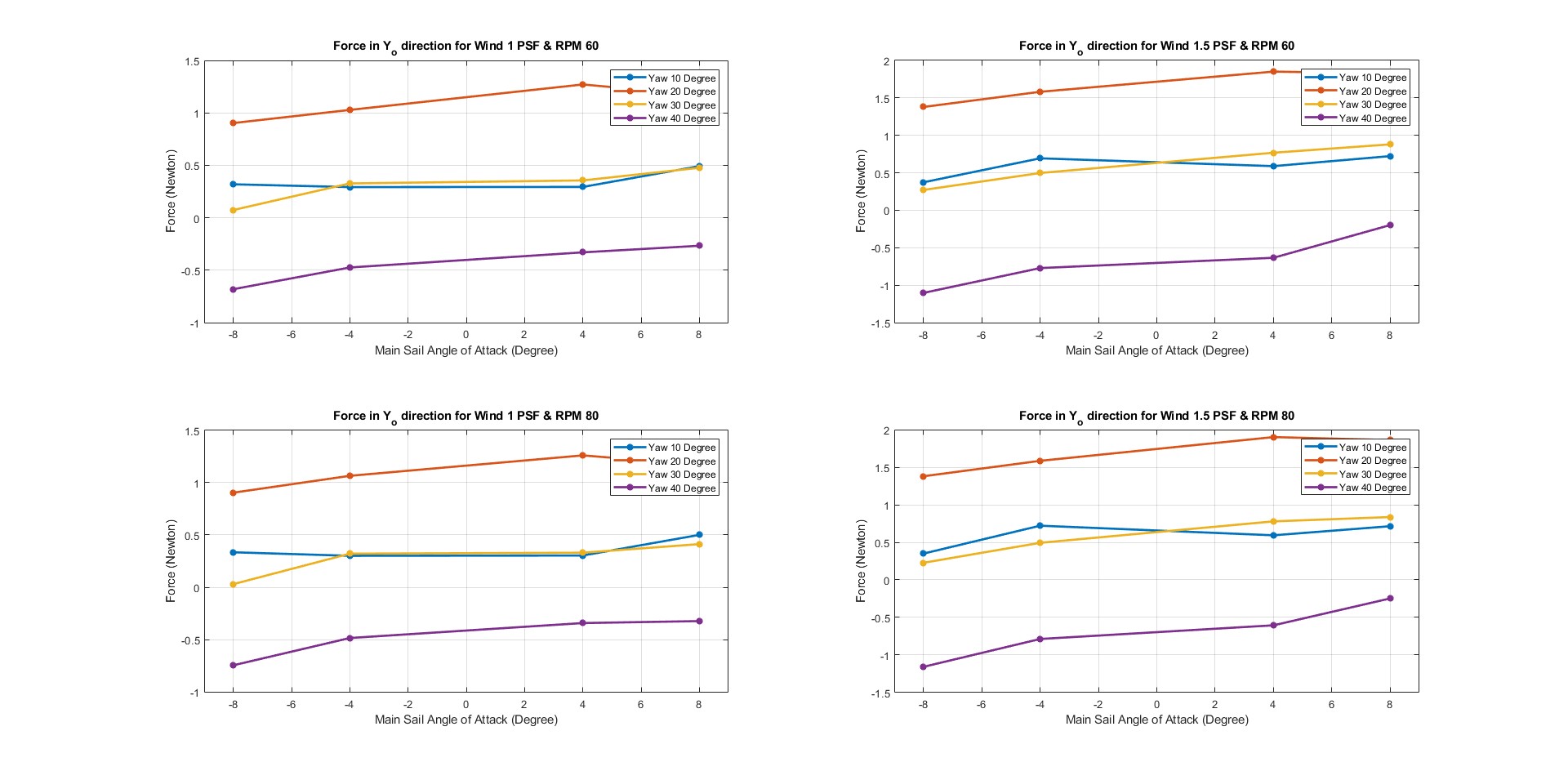}
    \caption{Compiled Plot: Lateral ($F_y$) force (i.e. Lift) data for main sail angle sweep.}
    \label{fig:results_compiled_fy}
\end{figure}

\begin{figure}[H]
    \centering
    \includegraphics[width=0.95\linewidth]{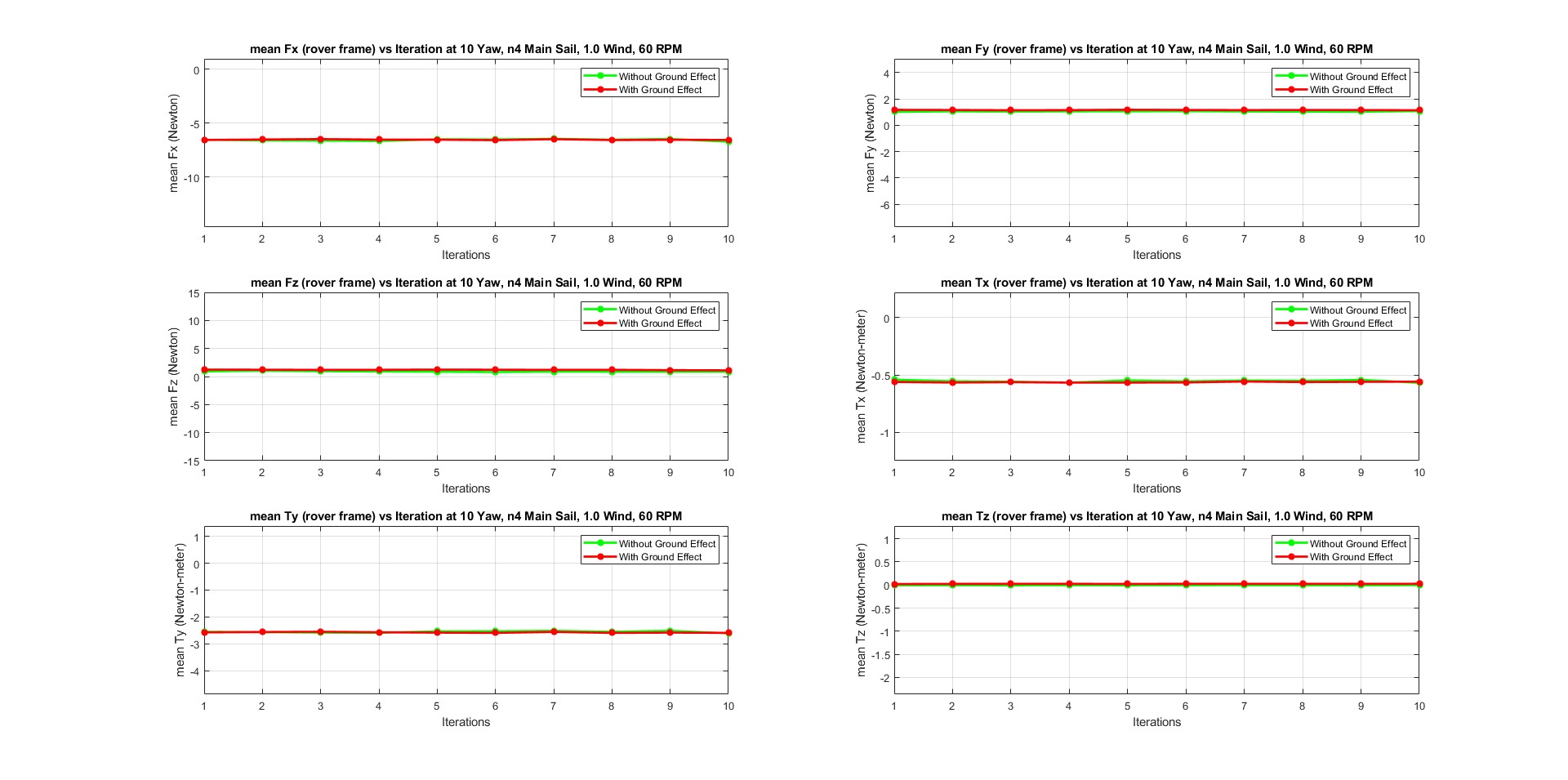}
    \caption{Comparison in forces and torques with \& without Ground effect for One Test Case }
    \label{fig: ground effect data}
\end{figure}

Above are the various plots under varying wind conditions, main sail angles of attack, yaw angles, and wheel RPM's. The measurements in the wind-aligned $X_0$ and lateral $Y_0$ directions representing drag and lift, respectively, reveal several key trends. 

In the $X_0$ direction (drag), forces were generally negative, as expected, and their magnitude increased with wind speed and yaw angle. The largest drag values were observed at Yaw = 30° and 40°, particularly under 1.5 PSF wind conditions. Changes in the main sail angle of attack had a moderate influence on drag, with slightly reduced drag near 0° to +8° AoA for certain configurations. However, the drag trend was relatively flat across angles for lower yaw values.

In the $Y_0$ direction (lift), the force increased more significantly with increasing sail angle of attack and was maximized for Yaw = 20°, especially at higher wind speeds and RPMs. The lift trends were consistent and favorable up to +8° AoA, with the peak lift reaching around 2 N. At Yaw = 40°, lift was reduced or even negative in some cases, indicating inefficient sail orientation at steep yaw angles. 

These results demonstrate the interdependent nature of yaw angle and sail orientation on aerodynamic performance. The observed trends will inform future dynamic modeling and controller co-design efforts aimed at maximizing net propulsion while minimizing energy loss from drag.

\vspace{12pt}
\section{\centering CONCLUSION}

This study presents a comprehensive wind tunnel experimental campaign to characterize the aerodynamic behavior of the Spherical Sailing Omnidirectional Rover (SSailOR), a novel sail-driven mobile platform. The custom-designed test rig enabled isolated force measurements across a test matrix of wind speeds, yaw angles, main sail angles of attack, and rover RPM's.

Force measurements in the wind-aligned direction $X_0$ revealed that drag magnitudes increased with yaw angle and wind speed, with the highest drag observed at yaw angles of 30° and 40° under 1.5 PSF wind conditions. Drag trends across sail angles of attack were relatively flat at lower yaw angles, suggesting that yaw-induced exposure dominates over sail orientation in affecting drag.

In the lateral direction $Y_0$, representing lift, forces were more sensitive to sail angle of attack. Lift increased consistently with angle of attack, reaching a maximum near +8°, especially for yaw angles of 10° to 20°. At 40° yaw, lift was significantly diminished or even negative, highlighting the importance of coordinated yaw-sail configuration for efficient propulsion.

These insights reinforce the coupled nature of sail actuation and yaw steering in determining SSailOR's mobility performance. The results will support the ongoing co-design of dynamic models and control strategies by providing experimental validation and helping identify optimal operating configurations. This work lays the foundation for future hardware-in-the-loop validation, real-world deployment, and energy-aware control of wind-powered spherical rovers.

\vspace{12pt}
\section{\centering ACKNOWLEDGMENTS}

The authors thank the Department of Mechanical and Aerospace Engineering at North Carolina State University and Dr. Andre P. Mazzoleni for their continued support and access to essential testing infrastructure. The authors also thank Dr. Chris Vermillion from the University of Michigan for his collaboration and guidance throughout the project. This work was supported in part by the National Science Foundation (NSF) under Project ID 2348361.

\vspace{12pt}


\end{document}